\documentclass[useAMS,usenatbib,fleqn]{mn2e}

\usepackage{graphicx}
\usepackage{float}
\usepackage{amsmath}
\usepackage{amssymb}
\usepackage[labelformat=empty,labelsep=none]{subcaption}
\usepackage{epsfig}
\usepackage{times}
\usepackage{multirow}
\usepackage{color}
\usepackage{gensymb}

\def\gtsim {\gtrsim}   
\def\ltsim {\lesssim}   

\newcommand{\dummytitle}[1]{}

\newcommand{\msun}{{\,\rm M}_\odot }
\newcommand{\atd}{ATLAS$^{\rm 3D}$ }

\defcitealias{pt14}{TK14}
\defcitealias{pt15a}{TK15a}
\defcitealias{pt15b}{TK15b}

\title[Kinematics of simulated galaxies]{The origin of kinematically distinct cores and misaligned gas discs in galaxies from cosmological simulations}
\author[P.~Taylor, C.~Federrath, and C.~Kobayashi]{Philip~Taylor$^1$\thanks{E-mail: philip.1.taylor@anu.edu.au}, Christoph~Federrath$^1$, and Chiaki~Kobayashi$^2$\\
$^1$Research School of Astronomy and Astrophysics, Australian National University, Canberra, ACT 2611, Australia\\
$^2$Centre for Astrophysics Research, School of Physics, Astronomy and Mathematics, University of Hertfordshire, AL10 9AB, UK}

\begin{document}
				
\date{Accepted  Received ; in original form}

\pagerange{\pageref{firstpage}--\pageref{lastpage}} \pubyear{}

\maketitle

\label{firstpage}

\begin{abstract}
Integral field spectroscopy surveys provide spatially resolved gas and stellar kinematics of galaxies.
They have unveiled a range of atypical kinematic phenomena, which require detailed modelling to understand.
We present results from a cosmological simulation that includes stellar and AGN feedback.
We find that the distribution of angles between the gas and stellar angular momenta of galaxies is not affected by projection effects.
We examine five galaxies ($\approx 6$ per cent of well resolved galaxies) that display atypical kinematics; two of the galaxies have kinematically distinct cores (KDC), while the other three have counter-rotating gas and stars.
{ All five form the majority of their stars in the field, subsequently falling into cosmological filaments where the relative orientation of the stellar angular momentum and the bulk gas flow leads to the formation of a counter-rotating gas disc.
The accreted gas exchanges angular momentum with pre-existing co-rotating gas causing it to fall to the centre of the galaxy.
This triggers low-level AGN feedback, which reduces star formation.
Later, two of the galaxies experience a minor merger (stellar mass ratio $\sim1/10$) with a galaxy on a retrograde orbit compared to the spin of the stellar component of the primary.
This produces the KDCs, and is a different mechanism than suggested by other works.}
The role of minor mergers in the kinematic evolution of galaxies may have been under-appreciated in the past, and large, high-resolution cosmological simulations will be necessary to gain a better understanding in this area.

\end{abstract}



\section{Introduction}
\label{sec:intro}

Galaxies may be broadly separated into two categories based on their morphology: spirals (late-type galaxies; LTGs) and ellipticals (early-type galaxies; ETGs).
Spiral galaxies are thin, rotationally supported discs with spiral arms, while ellipticals were traditionally believed to be dispersion-supported systems with smooth, featureless photometry.
However, the appearance of elliptical galaxies belies the potentially complex kinematics of their constituent gas and stars.

In recent years, integral-field spectroscopy (IFS) has superseded long-slit spectroscopy as the standard tool for analysing galaxy kinematics.
Several surveys of hundreds or thousands of galaxies now exist, including SAURON \citep{dezeeuw02}, \atd \citep{cappellari11}, CALIFA \citep{sanchez12}, SAMI \citep{croom12}, SLUGGS \citep{brodie14}, { MASSIVE \citep{ma14}}, S7 \citep{dopita15,thomas17}, and MaNGA \citep{yan16}, with the next generation of IFS surveys such as Hector \citep{Hector} set to observe 100,000 galaxies.
These surveys unveiled two kinematically distinct sub-classes of ETGs - fast rotators and slow rotators -  as well as a family of diverse kinematic sub-structures in ETGs (and LTGs to a lesser extent) including kinematically distinct cores (KDCs), misaligned or counter-rotating gas and stars, and so-called double-$\sigma$ (2$\sigma$) galaxies.
\citet{cappellari16} provides a comprehensive review of galaxy kinematics; we focus here on those aspects most important to this work.

Fast and slow rotators are defined by their position in the $\lambda_{R_{\rm e}}$ -- $\varepsilon$ plane, with $\varepsilon$ the observed ellipticity, and $\lambda_{R_{\rm e}}$ the angular momentum parameter \citep{emsellem07} given by
\begin{equation}
	\lambda_{R_{\rm e}} = \frac{\left<R\left|V\right|\right>}{\left<R\sqrt{V^2+\sigma^2}\right>},
\end{equation}
where $\left<\cdot\right>$ indicates a flux-weighted average over spaxels, and all spaxels within one effective radius are considered.
Slow rotators tend to have smaller $\lambda_{R_{\rm e}}$ and $\varepsilon$.
They are found in dense environments and have likely experienced little \emph{in situ} star formation since high redshift, growing mainly through dry (gas-poor) mergers \citep[e.g.,][]{emsellem07,naab14}.
Fast rotators are more numerous, and are found outside of cluster environments.
Despite their elliptical morphology, all fast rotators contain a disc-like component \citep{krajnovic08}.

KDCs are seen in stellar velocity maps as a change in the position angle of the kinematic axis with radius.
They are seen in only a small fraction of galaxies, almost exclusively slow rotators; \citet{krajnovic08} estimated that around 29 per cent of ETGs in SAURON hosted a KDC, and \citet{krajnovic11} revised this to only 7 per cent based on \atd data.
Their origin has not yet been explained in detail, but much work has been undertaken to understand how they are formed.
\citet{jesseit07} used idealised N-body and N-body/SPH simulations of merging spiral galaxies to study how the mass ratio, gas fraction, and orbital parameters affected the presence of a KDC in the merger remnant, finding that most kinematically interesting systems form from equal mass mergers.
{ \citet{hoffman10} investigated the impact of gas fraction in binary spiral mergers in more detail using N-body/SPH simulations, with gas fractions of 15-20 per cent likely to produce a KDC.}
This result was echoed by a similar study by \citet{bois11} who found that most slow rotators that formed by the major merger of two disc galaxies host a KDC.
Similarly, the semi-analytic model (SAM) of \citet{khochfar11} concluded that KDCs form from wet (gas-rich) mergers at high redshift.
\citet{naab14} analysed zoom-in simulations, which included star formation and gas cooling, of 44 haloes taken from a low-resolution cosmological simulation, producing more realistic environments in which the galaxies resided. 
These simulations were able to qualitatively reproduce the kinematic features seen in IFS surveys, { but lacked strong feedback from active galactic nuclei (AGN) which can impact central kinematics \citep[e.g.,][]{dubois13} as well as the gas fraction and star formation histories of the simulated galaxies, especially at high mass \citep[e.g.,][]{pt17a}.}

Related to KDCs, 2$\sigma$ galaxies have two symmetrical peaks in their velocity dispersion maps along their major axis.
Such features are seen in only $\sim4$ per cent of \atd galaxies, mostly flattened systems viewed near edge on \citep{krajnovic11}.
They are thought to be due to two counter-rotating, co-spatial disc structures \citep{krajnovic11}, and may be formed by the merger of two disc galaxies with opposite spins \citep[e.g.,][]{bois11} or the accretion of extra-galactic gas \citep[e.g.,][]{coccato11,algorry14}.

Misaligned gas and stellar kinematics are seen in a significant minority of ETGs.
\citet{davis11} found $36\pm5$ per cent of ETGs from \atd were more than $30\degree$ misaligned.
More recently, Bryant et al. (in prep.) found that $45\pm6$ per cent of ETGs from the SAMI survey showed similar misalignment.
Both studies suggest that mergers do not play a role in the creation of misaligned gas and stellar kinematics, but that accretion of extra-galactic gas is more important.

Understanding the formation and evolution of complex kinematic features is extremely challenging from observations alone, and numerical simulations are also needed.
Studies such as \citet{jesseit07} and \citet{bois11} used idealised simulations of the merger of two galaxies to examine the resulting kinematics.
The advantage of this method is that parameters such as mass ratio, gas fraction, and orbital parameters can be varied systematically in the initial conditions to assess their relative influence.
It is also possible to run these simulations at relatively high mass and high spatial resolution.
However, they do not capture true cosmological environment in which the galaxies evolve, and so neglect minor mergers and gas flows along filaments.
\citet{naab14} performed cosmological zoom-in simulations in which dark matter haloes were selected from a low-resolution cosmological simulation, and resimulated at higher resolution including gas and stellar physics.
In this way the galaxies could be evolved in a cosmological context, but AGN feedback was not included.

In this paper, we use a cosmological, chemodynamical simulation to investigate the origin of kinematic irregularities seen in a handful of our simulated galaxies.
In Section \ref{sec:sims} we describe the simulation in detail, as well as our analysis procedure.
Section \ref{sec:gals} introduces the galaxies with atypical kinematics, and their formation and evolution is examined in detail in Section \ref{sec:results}.
Finally, we present our conclusions in Section \ref{sec:conc}.

\section{Simulation Setup and Analysis Procedure}
\label{sec:sims}

\subsection{The Simulation}

The simulation used in this paper is a cosmological, chemodynamical simulation, which was introduced in \citet{pt15a}.
Our simulation code is based on the smoothed particle hydrodynamics (SPH) code {\sc gadget-3} \citep{springel05gadget}, updated to include:  star formation \citep{ck07}, energy feedback and chemical enrichment from supernovae \citep[SNe II, Ibc, and Ia,][]{ck04,ck09} and hypernovae \citep{ck06,ck11a}, and asymptotic giant branch (AGB) stars \citep{ck11b}; heating from a uniform, evolving UV background \citep{haardt96}; metallicity-dependent radiative gas cooling \citep{sutherland93}; and a model for black hole (BH) formation, growth, and feedback \citep{pt14}, described in more detail below.
We use the initial mass function (IMF) of stars from \citet{kroupa08} in the range $0.01-120\msun$, with an upper mass limit for core-collapse supernovae of $50\msun$.

The initial conditions for the simulation consist of $240^3$ gas and dark matter particles in a periodic, cubic box $25\,h^{-1}$ Mpc on a side, giving spatial and mass resolutions of $1.125\,h^{-1}$ kpc and $M_{\rm DM}=7.3\times10^7\,h^{-1}\msun$, $M_{\rm gas}=1.4\times10^7\,h^{-1}\msun$, respectively.
This resolution is sufficient to resolve distinct kinematic features in massive galaxies.
We employ a WMAP-9 $\Lambda$CDM cosmology \citep{wmap9} with $h=0.7$, $\Omega_{\rm m}=0.28$, $\Omega_\Lambda=0.72$, $\Omega_{\rm b}=0.046$, and $\sigma_8=0.82$.

Our AGN model is unique; BHs form from gas particles that are metal-free and denser than a specified critical density, mimicking the most likely formation channels in the early Universe as the remnant of Population {\sc III} stars \citep[e.g.,][]{madau01,bromm02,schneider02} or via direct collapse of a massive gas cloud \citep[e.g.,][]{bromm03,koushiappas04,agarwal12,becerra15,regan16a,hosokawa16}.
The BHs grow through Eddington-limited Bondi-Hoyle gas accretion and mergers.
Two BHs merge if their separation is less than the gravitational softening length and their relative speed is less than the local sound speed.
A fraction of the energy liberated by gas accretion is coupled to neighbouring gas particles in a purely thermal form.
 
In previous works, we have compared the simulation used in this paper with another having the same initial conditions, but without the inclusion of any BH physics.
We showed that the inclusion of AGN feedback leads to simulated galaxies whose properties more closely match those of observed galaxies, both at the present day and high redshift \citep{pt15a,pt15b,pt16,pt17b}, and quantified the effects of AGN feedback on the host galaxy and its immediate environment \citep{pt15b,pt17a}.

\subsection{Angular Momentum}\label{sec:}

Angular momentum is the most useful quantity for understanding the kinematics of our simulated galaxies.
For each particle,
\begin{equation}
	\mathbf{j} = m\left(\mathbf{r}-\mathbf{r}_{\rm com}\right)\times\left(\mathbf{v}-\mathbf{v}_{\rm com}\right),
\end{equation}
where $m$ is the particle mass, $\mathbf{r}$ its position, $\mathbf{v}$ its velocity, and $\mathbf{r}_{\rm com}$ and $\mathbf{v}_{\rm com}$ are the position and velocity of the galaxy centre of mass, respectively.
The total angular momentum is then 
\begin{equation}
	\mathbf{J} = \sum_{i}{\mathbf{j}_{i}},
\end{equation}
where the sum may run over all particles in a galaxy, or subsets such as stars or gas.

\begin{figure*}
	\centering
	\includegraphics[width=0.98\textwidth,keepaspectratio]{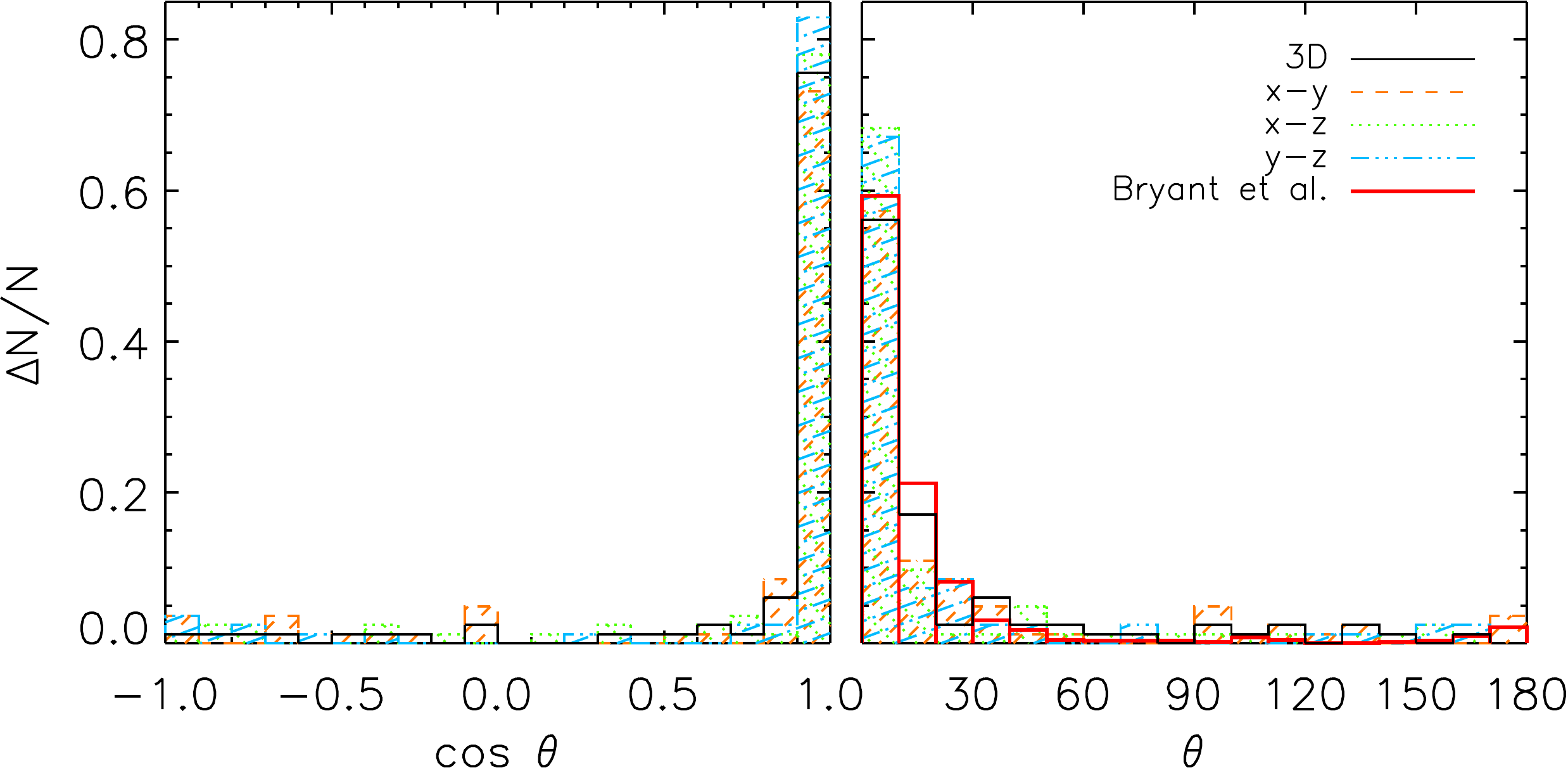}
	\caption{\emph{Left panel}: Distributions of kinematic misalignment between gas and stars ($\cos\theta$, equation \eqref{eq:costheta}) from 3D angular momenta and projections in our simulated galaxies.
	\emph{Right panel}: Distributions of $\theta$ from simulated galaxies (black) and observations (thick, solid line; Bryant et al. (in prep.)).}
	\label{fig:costheta}
\end{figure*}

It is instructive to measure the angle between the angular momenta of the stars and gas in our simulated galaxies since this gives a simple indication of kinematic misalignment, and can be directly compared to observational data.
This angle is given by
\begin{equation}\label{eq:costheta}
	\cos\theta = \mathbf{\hat J}_{*}\cdot\mathbf{\hat J}_{\rm gas},
\end{equation}
where $\mathbf{\hat J}$ denotes the normalised angular momentum such that $\mathbf{\hat J}\cdot\mathbf{\hat J}=1$.
In our simulation it is possible to measure the three-dimensional (3D) vectors $\mathbf{\hat J}_{*}$ and $\mathbf{\hat J}_{\rm gas}$, whereas IFU observations can only obtain line-of-sight velocity profiles for gas and stars, from which $\theta$ can be measured.
To compare to observations, taking into account this projection effect, we define $\theta_{ab}$ as
\begin{equation}
	\cos\theta_{ab} = \frac{J_{*,a}J_{{\rm gas},a} + J_{*,b}J_{{\rm gas},b}}{\sqrt{\left(J_{*,a}^2+J_{*,b}^2\right)\left(J_{{\rm gas},a}^2 + J_{{\rm gas},b}^2\right)}},
\end{equation}
with $ab$ one of $xy$, $xz$, and $yz$.
The distributions of $\cos\theta$, $\cos\theta_{xy}$, $\cos\theta_{xz}$, and $\cos\theta_{yz}$ are shown in Fig. \ref{fig:costheta}.
The majority of galaxies have aligned ($\cos\theta=1$) gas and stellar kinematics.
A small fraction of galaxies display anti-alignment ($\cos\theta=-1$) in which the gas rotates in the same plane as the stars, but in the opposite direction.
The distributions of $\cos\theta_{ab}$ are consistent with one-another and $\cos\theta$, indicating that the projection effect does not statistically alter observational measurements of this quantity.

In the right-hand panel of Fig. \ref{fig:costheta}, we show the distributions of $\theta$ of our 82 well resolved galaxies, as well as the observed distribution from Bryant et al. (in prep.) in the SAMI survey.
There is excellent qualitative agreement between the simulated and observed distributions, however the simulation over-produces galaxies with intermediate misalignment ($60\degree\ltsim\theta\ltsim150\degree$) compared to observations.
{ Bryant et al. (in prep.) concluded that there is a statistically significant secondary peak at $\theta=180\degree$ in the SAMI data (thick, solid line), which is not reproduced by the simulated galaxies, whose distribution of $\theta$ is consistent with being uniform for $\theta\gtsim45\degree$.
This may be due to the limited sample size (82 well resolved galaxies) afforded by the simulation.}

\subsection{Kinematic Maps}\label{sec:}

IFU observations of galaxies provide individual spectra across the face of the galaxy.
By identifying and fitting emission and absorption lines (for gas and stars, respectively) in each spectrum, velocity maps can be constructed.
Of course, in our simulation, the velocity of all particles is available.
To generate kinematic maps, we use the following method.

Galaxies are rotated so that the net angular momentum of their stars lies along the $z$-axis (i.e., along the line of sight)\footnote{Note that this means we do not expect to find 2$\sigma$ galaxies.}.
We smooth the properties of individual particles over many pixels on a regular grid.
Specifically, the average velocity in the $j^{\rm th}$ pixel is given by
\begin{equation}
	\left<\mathbf{v}\right>_j = \frac{\sum_i \left(\mathbf{v}_i - \mathbf{v}_{\rm com}\right)f_{ij} w_i}{\sum_i f_{ij} w_i},
\end{equation}
where the sum is over all particles of interest, typically the gas or stars of a galaxy, and $f_{ij}$ denotes the fractional contribution of particle $i$ to pixel $j$:
\begin{equation}
	f_{ij} = \left(2\pi (0.3h)^2\right)^{-1} \int {\rm d}x\,{\rm d}y \exp\left(-\frac{\left(x-x_i\right)^2+\left(y-y_i\right)^2}{2\left(0.3h\right)^2}\right),
\end{equation}
where $h$ is the gravitational softening length of the simulation, and the factor 0.3 enters to closely match the spline kernel used in the simulation.
The integral is evaluated over the area of the pixel.
The weights $w_i$ are either particle mass, or, for closer comparison to observational data, $V$-band luminosity ($L_V$) for stars and star formation rate (SFR) for gas.
Maps generated weighting by $L_V$ or SFR are quantitatively similar to mass-weighted maps, but tend to be noisier, and some are qualitatively different; we focus on mass-weighted maps in the following sections, but see Appendix \ref{sec:app:weight} for maps with other weightings, { and Section \ref{sec:observe} for a discussion on the observability of these features.}

To compare more closely to the analysis of IFU data, we generate maps of estimated signal-to-noise ratio (S/N) which we Voronoi bin to some target S/N \citep{cappellari03}.
We use the Wide-Field Spectrograph (WiFeS) performance calculator\footnote{http://www.mso.anu.edu.au/rsaa/observing/wifes/performance.shtml}, appropriate for the S7 survey\footnote{{The {\emph Siding Spring Southern Seyfert Spectroscopic Snapshot Survey} \citep{dopita15,thomas17}.}}, to estimate S/N given the $V$-band surface brightness of each pixel, and assuming a 1800s integration time.
Note that this method differs slightly from observations, where the binning achieves a target S/N of spectral features, rather than total brightness.
The bins so produced are applied to both the stellar and gas kinematic maps; in observations, stellar spectral features are typically much weaker than emission lines from gas, and so the gas emission lines in each bin automatically reach the target S/N.
Throughout we adopt a target S/N of 25, but see Appendix \ref{sec:app:sn} for the effect of this value on the kinematic maps, as well as the un-binned maps.

\begin{figure*}
	\centering
	\includegraphics[width=0.98\textwidth,keepaspectratio]{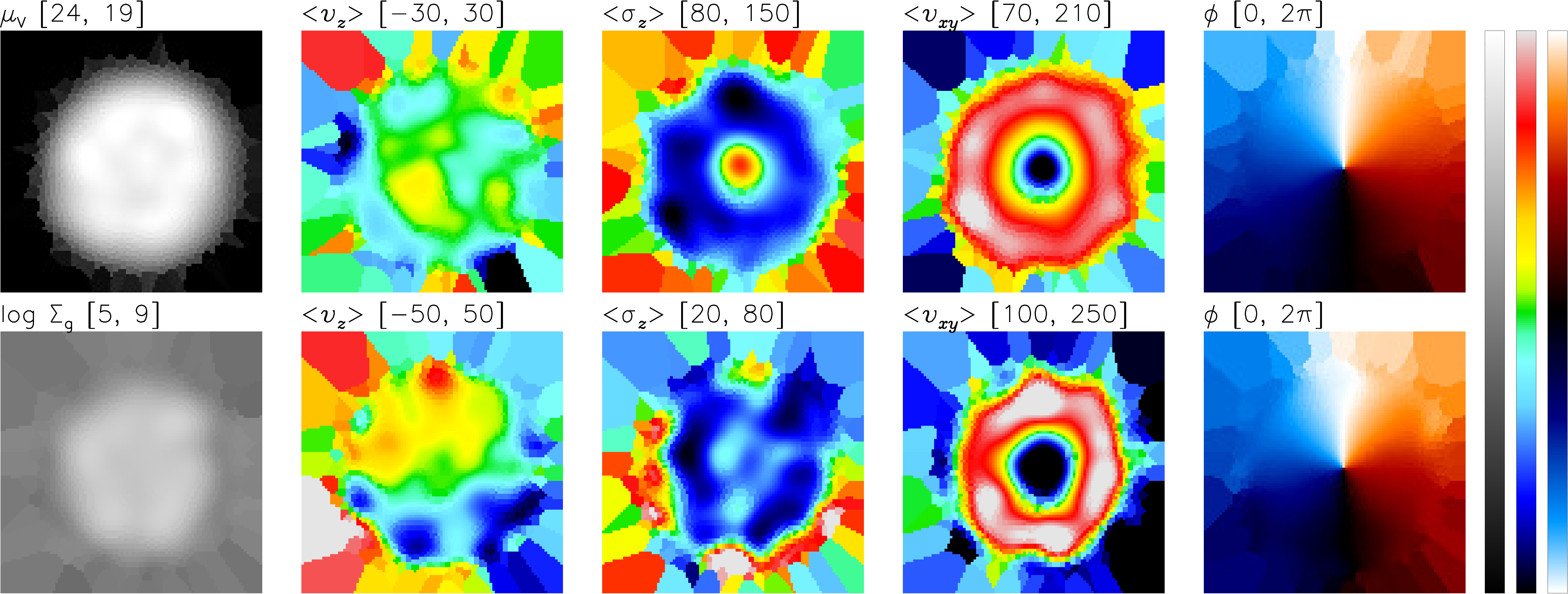}
	\caption{Kinematic maps for ga0008, a prototypical disc galaxy from our simulation.
	The top row shows stellar properties, and the bottom row is for gas.
	The first column shows $V$-band surface brightness (mag\,arcsec$^{-2}$) and gas surface density (M$_\odot$\,kpc$^{-2}$), the second average line-of-sight velocity $\left<v_z\right>$ (km\,s$^{-1}$), the third line-of-sight velocity dispersion $\sigma_z$ (km\,s$^{-1}$), the fourth $v_{xy}=\sqrt{\left<v_x\right>^2+\left<v_y\right>^2}$ (km\,s$^{-1}$), and the fifth $\phi=\tan^{-1}\left(\left<v_y\right>/\left<v_x\right>\right)$.
	The range of values in each panel is shown in brackets.
	Panels are 20 kpc ($\approx6.7R_{\rm e}$) on a side.}
	\label{fig:ga0008}
\end{figure*}
We estimate the line-of-sight velocity dispersion as 
\begin{equation}
	{\sigma_z^2}_j = \frac{\left<v_z^2\right>_j - \left<v_z\right>_j^2}{1 - \sum_i \left(f_{ij}w_i\right)^2 / \left(\sum_i f_{ij}w_i\right)^2}.
\end{equation}
In addition, we have access to more information than observers, and the quantity $\phi = \tan^{-1}\left(\left<v_y\right> / \left<v_x\right>\right)$, describing the direction of motion in the $x$--$y$ plane, is found to be useful.
To give a clear example, Fig. \ref{fig:ga0008} shows maps of surface brightness and gas surface density, as well as $\left<v_z\right>$, $\sigma_z$, $v_{xy} = \sqrt{\left<v_x\right>^2+\left<v_y\right>^2}$, and $\phi$ for the stars (top row) and gas (bottom row) of galaxy ga0008.
Each panel is 20 kpc ($\approx6.7R_{\rm e}$) on a side.
ga0008 is a disc galaxy in which the stars and gas co-rotate in the same plane ($\cos\theta=0.998$).

After rotating the galaxy as described above, $\left<v_z\right>$ is close to 0 across the disc (second column), and $\sigma_z$ is larger in magnitude (third column).
The speed of both gas and stars in the plane, $v_{xy}$, increases radially from the centre of the galaxy, reaching a broad peak around 8 kpc ($\approx2.8R_{\rm e}$, where $R_{\rm e}$ is the effective radius; see { Section 3.2 of} \citet{pt15a} for details) from the centre (fourth column).
In the final column of Fig. \ref{fig:ga0008}, we show maps of $\phi$ for the stars and gas, which indicates the direction of motion in the plane.
Both move clockwise, and variation in the value of $\phi$ is almost entirely azimuthal, as expected for a rotating disc.

\section{Galaxies Displaying Atypical Kinematics}\label{sec:gals}

\begin{table}
\caption{Present-day properties of the galaxies presented in Section \ref{sec:gals}.
Stellar mass, gas mass, BH mass, $5^{\rm th}$-nearest neighbour distance, and $\cos\theta$ (see equation \eqref{eq:costheta}) are given.}
	\begin{tabular}[width=0.5\textwidth]{ccccccc}
		Galaxy & $\log M_*$ & $\log M_{\rm gas}$ & $\log M_{\rm BH}$ & $R_{\rm e}$ & $s_5$ & $\cos \theta$ \\
		& $[\msun]$ & $[\msun]$ & $[\msun]$ & [kpc] & [kpc] &\\
		\hline 
		ga0045 & 10.8 & 9.8 & 6.2 & 3.2 & 273 & -0.84 \\
		ga0064 & 10.7 & 9.7 & 5.7 & 2.6 & 2,057 & -0.94 \\
		ga0074 & 10.6 & 9.9 & 6.1 & 2.3 & 957 & 0.91 \\
		ga0091 & 10.5 & 9.8 & 6.0 & 2.4 & 2,434 & 0.81 \\
		ga0099 & 10.5 & 9.9 & 5.9 & 2.0 & 1,651 & -0.75 
	\end{tabular}
\label{tab:gals}
\end{table}

Galaxies are identified using a parallel Friends-of-Friends (FoF) finder (based on a serial version provided by V.~Springel).
The code associates dark matter particles, separated by at most 0.02 times the mean inter-particle separation, into groups.
Gas, star, and BH particles are then joined to the group of their nearest dark matter neighbour.
Note that the `linking length' of 0.02 used here is smaller than typically adopted in the literature \citep[e.g., 0.2 in ][]{ck07}.
In these works, sub-halos are separated from the main FoF groups separately, whereas we use the smaller linking length to achieve the same result.

\begin{figure*}
	\centering
	\includegraphics[totalheight=0.98\textheight,keepaspectratio]{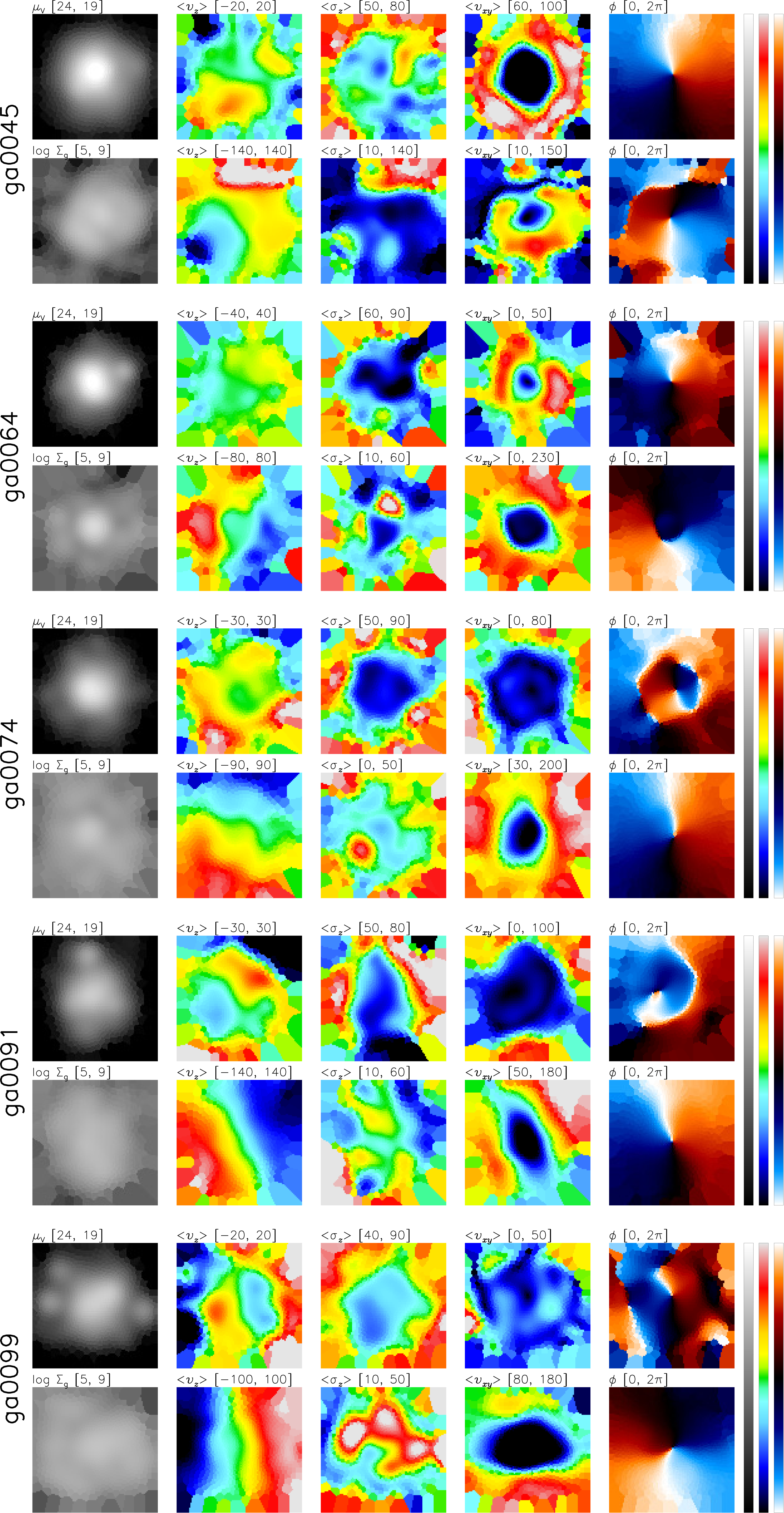}
	\caption{Kinematic properties of galaxies with atypical kinematics.
	Panels are $6R_{\rm e}$ on a side, but otherwise the same as in Fig. \ref{fig:ga0008} for each galaxy.}
	\label{fig:gaAll}
\end{figure*}

We generated maps of kinematic properties ($v_z$, $\sigma_z$, $v_x$, $v_y$, $v_{xy}$, $\phi$) for all galaxies identified by the FoF code that contained at least 1,000 star particles and 300 gas particles, corresponding to $\sim 6\times 10^9\msun$ in each of gas and stars.
From these 82 galaxies ($10^{10}<M_*/\msun<6\times10^{11}$), we identified five with atypical kinematics.
These galaxies are denoted ga0045, ga0064, ga0074, ga0091, and ga0099.
Kinematic maps of these galaxies are shown in Fig. \ref{fig:gaAll}, their present-day properties are given in Table \ref{tab:gals}.
Note that all five galaxies have similar masses; this is due to the trade-off between requiring the galaxies to be sufficiently well resolved, and the scarcity of very massive galaxies in our $\left(25\,h^{-1}\,{\rm Mpc}\right)^3$ simulation box.
In calculating { the $5^{\rm th}$-nearest neighbour distance,} $s_5$, galaxies of all masses are considered, and the value depends on the number of nearby satellite galaxies (see \citealt{pt17b} for the details).
We describe each galaxy in detail below.

Galaxy ga0045 has a disc-like morphology\footnote{Due to the finite simulation resolution, accurate morphologies are often difficult to determine. In this paper, the quoted morphology is based on visual inspection of the face-on and edge-on surface brightness profile of each galaxy.}, and is the most massive member of a small group of galaxies, with $s_5=273$ kpc.
Nevertheless, its merging history is relatively quiet, with no mergers in the previous 9 Gyr, though it experiences a close ($<10$ kpc separation) flyby with a galaxy of stellar mass $\log M_*=9.4$ approximately 150 Myr before present.
Stellar kinematics of ga0045 are shown in the top row of Fig. \ref{fig:gaAll}; they display similar features to the galaxy ga0008, shown in Fig. \ref{fig:ga0008}.
The gas kinematics on the lower row do not; $\left<v_z\right>$ shows large variation in value and a clear gradient across the galaxy, and its magnitude tends to be greater than $\sigma_z$.
This is reflective of the fact that $\left|\cos\theta\right|\ne1$ (Table \ref{tab:gals}).
In addition, the final panel of the lower row, which shows $\phi$ for the gas, indicates that the gas in the central region rotates in an opposite sense to the stars.

The morphology of ga0064 is disc-like, it is a field galaxy with no close companions, and has $s_5=2.1$ Mpc.
Its merger history is quiet, having no mergers in the previous 7 Gyr, and no major mergers in its lifetime.
In the kinematic maps of ga0064 (Fig. \ref{fig:gaAll}), stars show low $\left<v_z\right>$ (second panel) and ordered rotation throughout most of the galaxy (fifth panel), but their speeds in the $x$-$y$ plane (fourth panel) are not azimuthally symmetric, as in ga0008.
Gas shows greater range and a clear gradient in $\left<v_z\right>$ (second panel of lower row), as well as very high $v_{xy}$ (fourth panel).
The gas rotates in the opposite sense to the stars across most of the galaxy (fifth panel), and is disturbed at the centre.

Galaxy ga0074 is a disturbed disc-like galaxy that is interacting with a low-mass ($\log M_*=9.0$) galaxy approximately 18 kpc from its centre, and has $s_5=957$ kpc.
It experiences several minor mergers with stellar mass ratios $\sim 1/10$.
Fig. \ref{fig:gaAll} shows kinematic maps for ga0074.
For the stars, the central regions of the galaxy show no gradient in $\left<v_z\right>$, coincident with low $\sigma_z$, low $v_{xy}$, and a counter-rotating stellar core (fourth panel).
There is evidence of a gradient in stellar $\left<v_z\right>$ in the outskirts of the galaxy, which is mirrored in the gas $\left<v_z\right>$.
In the plane, the gas displays typical disc-like behaviour, rotating in the same sense as the stars of the outer regions of the galaxy.

Galaxy ga0091 is an isolated elliptical galaxy.
It undergoes a merger with stellar mass ratio 1:5 at $z=0.13$ (1.7 Gyr before present), but otherwise evolves quiescently.
In Fig. \ref{fig:gaAll} it displays similar properties to ga0074, having a counter-rotating stellar core (fifth panel of upper row), regular gas rotation (fifth panel of lower row), and a gradient in $\left<v_z\right>$ for gas across the face of the galaxy (second panel of bottom row), reflecting the fact that $\cos\theta\ne1$ for this galaxy (Table \ref{tab:gals}).

Galaxy ga0099 is an elliptical galaxy, and the second-most massive member of a small group of galaxies, but otherwise isolated, accounting for its relatively large $s_5$ in Table \ref{tab:gals}.
It experiences mergers of stellar mass ratios 1:5.7 and 1:4.2 at $z=0.7$ and $z=1.0$, respectively.
The kinematic maps of Fig. \ref{fig:gaAll} show that both $\left<v_z\right>$ and $v_{xy}$ are low for stars (second and fourth panels of top row) compared to the random motions (third panel).
In the final panel of the top row, showing $\phi$ for the stars, there is rotation in the central region, but the outskirts of the galaxy show no ordered motion.
By contrast, the gas particles (bottom row) show clear rotation in the opposite direction to the central stars.
Additionally, they have a clear gradient in $\left<v_z\right>$, with speeds of larger magnitude than their random motions.

\section{The Origin of Kinematic Misalignments}\label{sec:results}

Of the five galaxies identified in the previous section, three have counter-rotating gas discs (ga0045, ga0064, and ga0099), while the others have KDCs (ga0074 and ga0091).
We focus on the counter-rotating gas discs first, examine the stellar KDCs in Section \ref{sec:kdc}, and discuss the observability of such kinematic features in Section \ref{sec:observe}.

\subsection{Counter-Rotating Gas Discs}
\label{sec:crgd}

\begin{figure*}
	\centering
	\includegraphics[width=\textwidth,keepaspectratio]{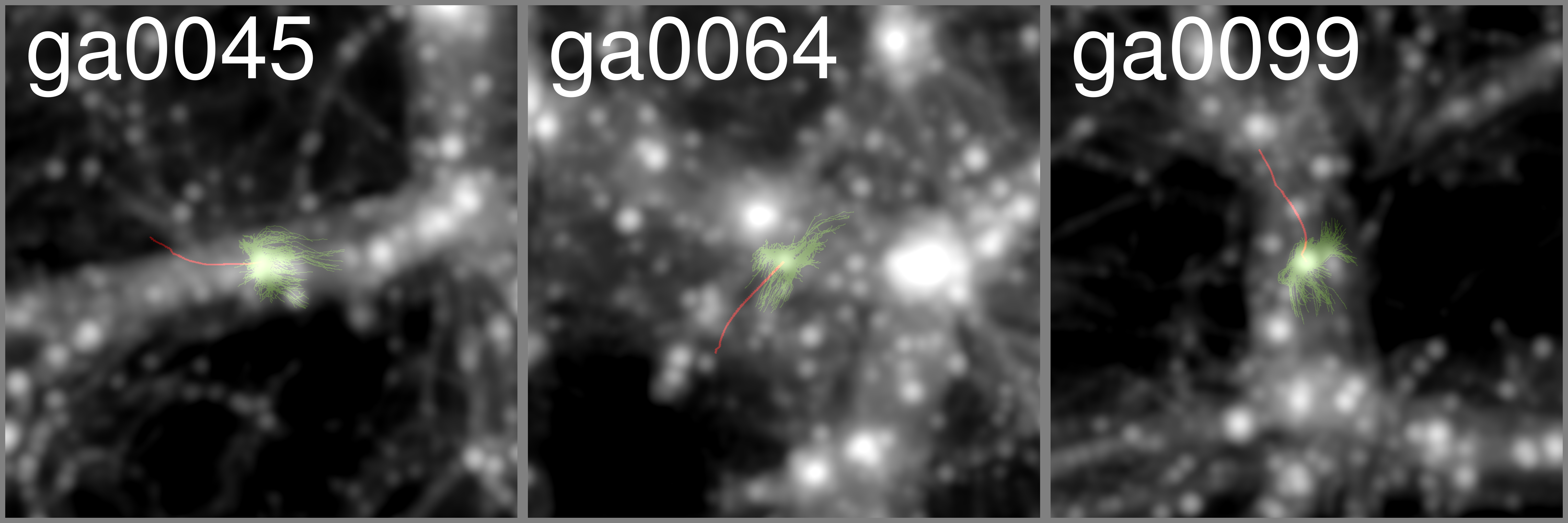}
	\caption{Maps of dark matter surface density in a region $20\times20\times6$ Mpc$^3$ centred on ga0045, ga0064, and ga0099.
	Galaxies are viewed such that $\mathbf{J}_*$ is perpendicular to the plane of the page.
	Overlaid in red is the path of each galaxy over cosmic time, and in green are the tracks of all gas particles that exist in the galaxy at $z=0$.}
	\label{fig:dm}
\end{figure*}

\begin{figure}
	\centering
	\includegraphics[width=0.48\textwidth,keepaspectratio]{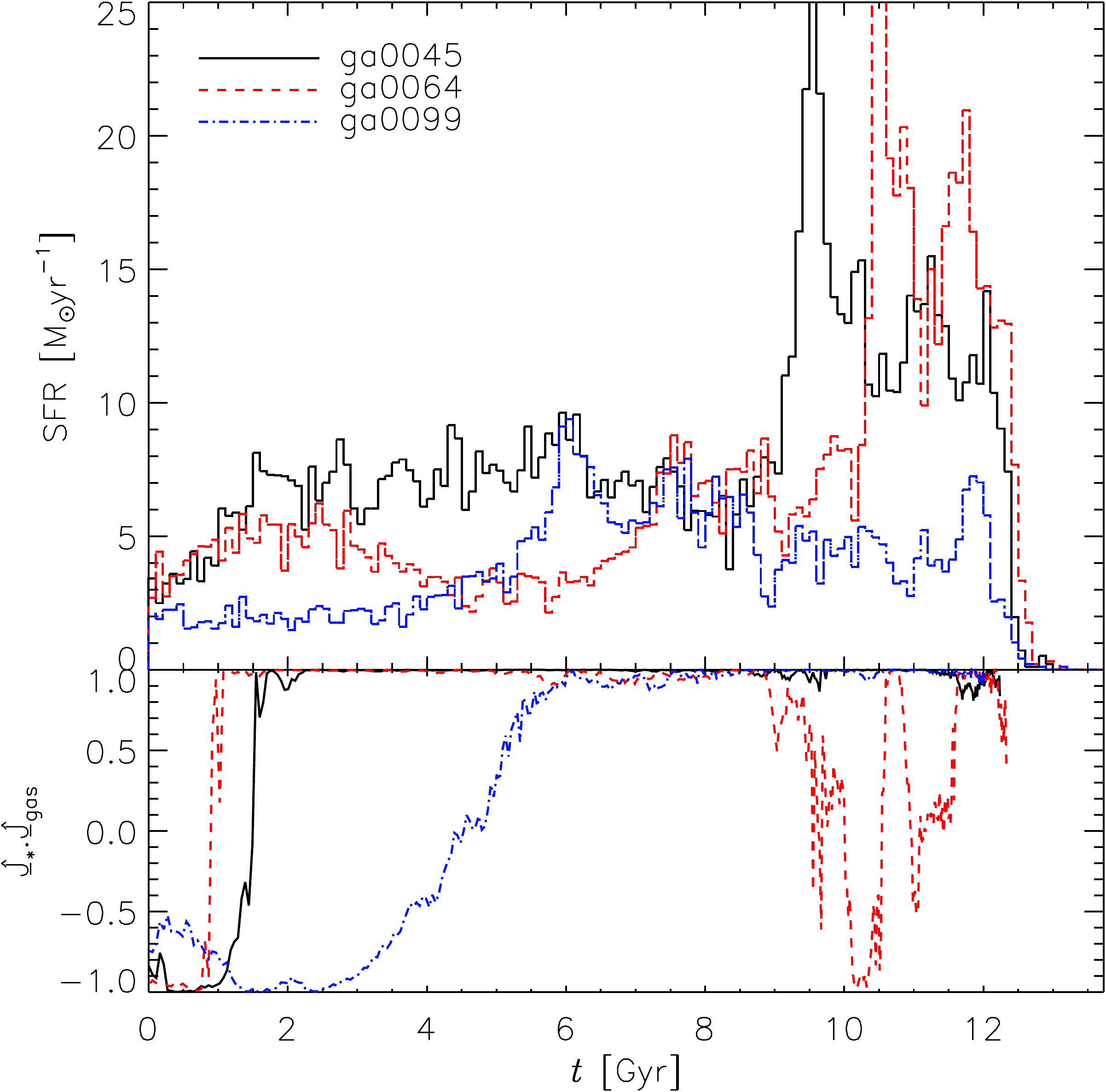}
	\caption{\emph{Top panel}: star formation histories for ga0045 (solid black line), ga0064 (dashed red line), and ga0099 (dot-dashed blue line).
	\emph{Lower panel}: evolution of $\hat{\mathbf{J}}_*\boldsymbol{\cdot}\hat{\mathbf{J}}_{\rm gas}$ for the same galaxies.}
	\label{fig:sfh_jsjg}
\end{figure}

Galaxies ga0045, ga0064, and ga0099 contain gas that rotates in an opposite sense to their stars; see the kinematic maps of Section \ref{sec:gals} for a visual representation.
{ Fig. \ref{fig:dm} shows the dark matter distribution in the vicinity of these galaxies, as well as their motion from high redshift ($z\sim4$) to the present (red lines).
All three form in the field away from massive dark matter filaments, and with gas and stellar angular momenta determined by their formation environment (note that the apparent path of ga0099 along the filament is a projection effect).
At later times, the galaxies fall into a filament, where the bulk gas motion is uncorrelated with the initial angular momentum of the galaxies.
In these galaxies, the relative orientation of the angular momentum of the galactic gas and the motion of gas accreted from the filaments leads to the formation of counter-rotating gas discs.
The accreted gas exchanges angular momentum with the galactic gas, causing it to fall towards the centre of the galaxy; subsequent accretion feeds the gas disc which rotates in the opposite direction to the stars.
}

{
The green tracks in Fig. \ref{fig:dm} show the motion over cosmic time of all gas particles that exist in the galaxies at the present day.
All of the present-day gas has been accreted from the filaments -- creating the counter-rotating gas disc -- while gas accreted earlier in the galaxy's evolution has been consumed by star formation, black hole accretion, or lost from the galaxy entirely.
Existing in the filaments, but away from nodes in the cosmic web, allows a large amount of gas to be accreted, while minimising the probability that the galaxy experiences a major merger that could disrupt the kinematics.
\citet{jin16} investigated galaxies with misaligned gas and stellar kinematics in the MaNGA survey, and also concluded that such galaxies are found in low-density environments.
}

{
Fig. \ref{fig:sfh_jsjg} demonstrates some of these points more quantitatively.
It }shows the star formation rate history of these galaxies (top panel) and the evolution of $\hat{\mathbf{J}}_*\boldsymbol{\cdot}\hat{\mathbf{J}}_{\rm gas}$ (bottom panel), with $\hat{\mathbf{J}}_*\boldsymbol{\cdot}\hat{\mathbf{J}}_{\rm gas}=1$ indicating perfect alignment, and $\hat{\mathbf{J}}_*\boldsymbol{\cdot}\hat{\mathbf{J}}_{\rm gas}=-1$ perfect anti-alignment, between the gas and stars in the galaxy.
In ga0045 and ga0064 the transition from co-rotating to counter-rotating gas happens on a short timescale, $\sim 250$ Myr.
The transition takes longer in ga0099, $\sim 3$ Gyr, but the mechanism is the same for all of the galaxies, as described above.
{ 
The timescale for this transition likely depends on several factors including the misalignment angle between the angular momentum of the co-rotating gas and the bulk flow within the filament, the accretion rate onto the galaxy, and the mass of co-rotating gas.
}

From Fig. \ref{fig:sfh_jsjg} it can be seen that the onset of the transition from co-rotating to counter-rotating gas is associated with a decline in SFR in all three galaxies.
The accreted counter-rotating gas transfers angular momentum to the co-rotating gas, causing it to fall towards the centre of the galaxy and fuel low-level AGN feedback.
This AGN activity keeps the gas in the galaxies hot, suppressing star formation, but is not sufficient to drive winds and the galaxies remain within the scatter of the star formation main sequence.
\citet{osman17} also found lower SFR in their high-resolution simulations of disc galaxies with counter-rotating gas.
In their model, which does not include AGN feedback, the counter-rotating gas suppressed spiral arm formation, and the associated regions of dense gas from which stars form.

\subsection{Stellar KDCs}
\label{sec:kdc}

\begin{figure*}
\centering
\begin{subfigure}{0.49\textwidth}
	\includegraphics[width=\textwidth,keepaspectratio]{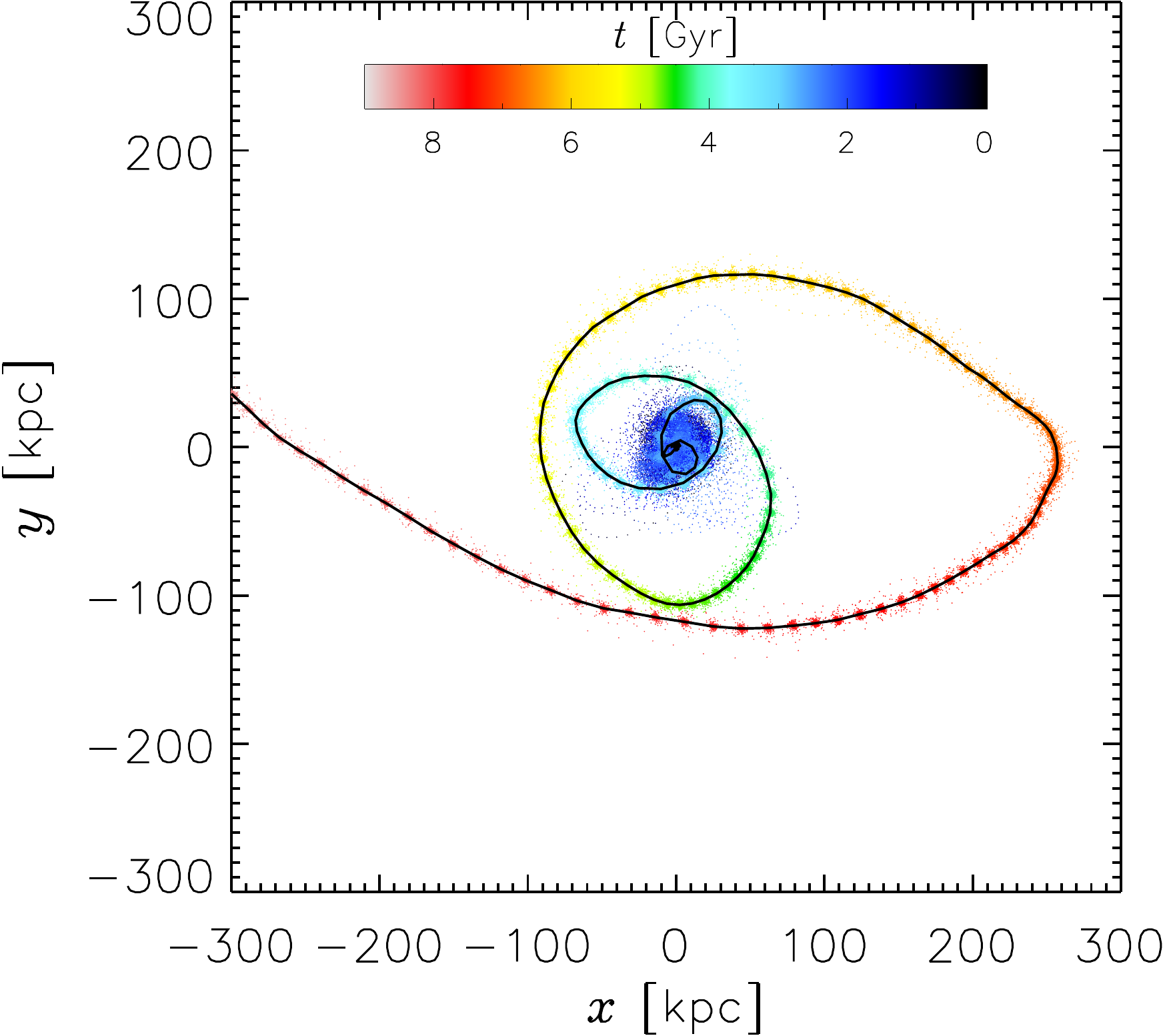}
	\caption{}
	\label{fig:0074merger}
\end{subfigure}
\begin{subfigure}{0.49\textwidth}
	\includegraphics[width=\textwidth,keepaspectratio]{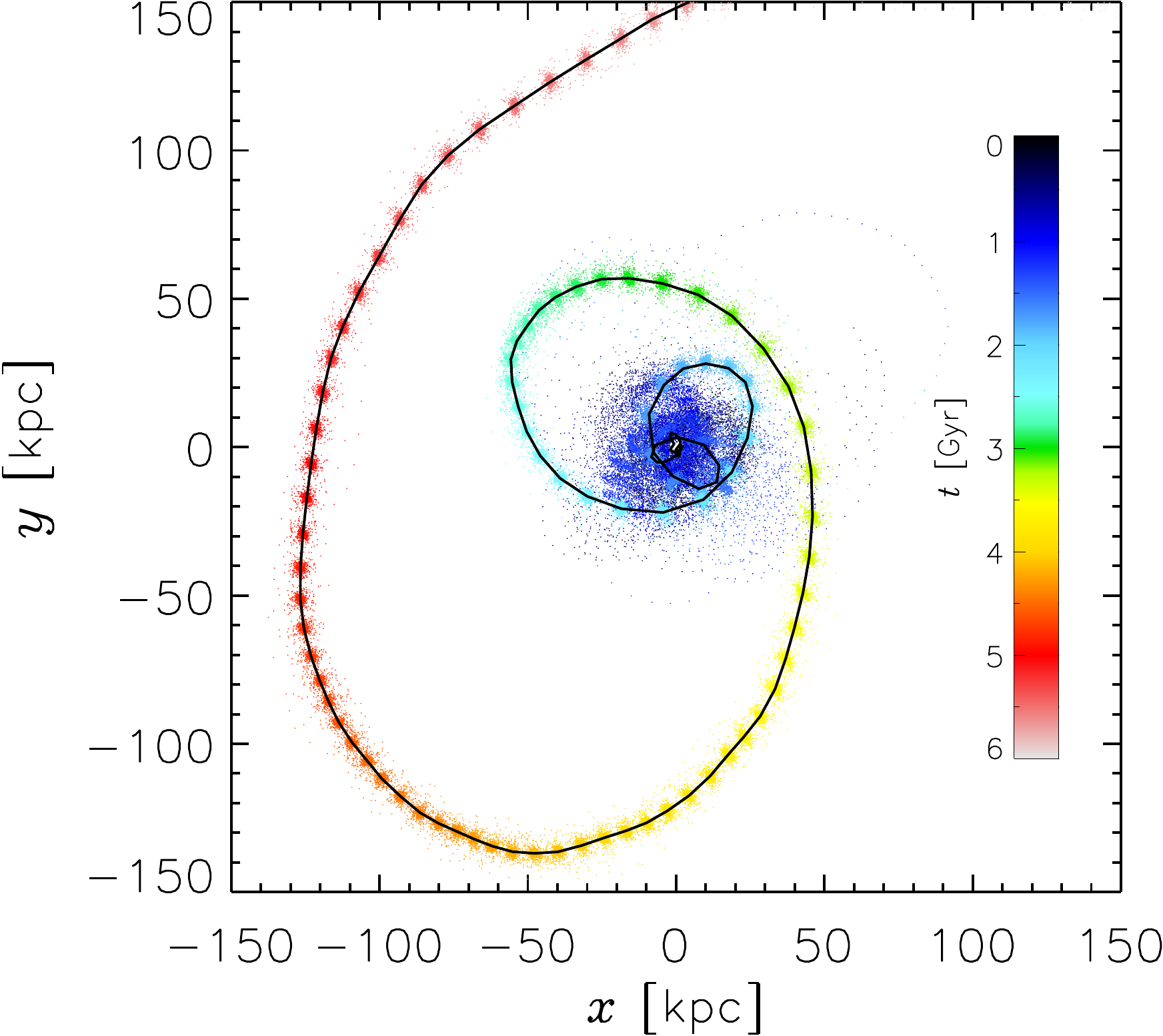}
	\caption{}
	\label{fig:0091merger}
\end{subfigure}
\caption{Positions of the galaxies that merge onto ga0074 (left panel; $R_{\rm e}=2.3$ kpc) and ga0091 (right panel; $R_{\rm e}=2.4$ kpc) at different lookback times.
	Particle locations are coloured by lookback time, and the solid black line traces the centre of mass of the secondary galaxy.
	The plane of the orbit is coincident with the plane of the page, { and the stellar component of the primary galaxies rotates clockwise}.}
\label{fig:orbits}
\end{figure*}

\begin{figure*}
	\centering
	\includegraphics[width=\textwidth,keepaspectratio]{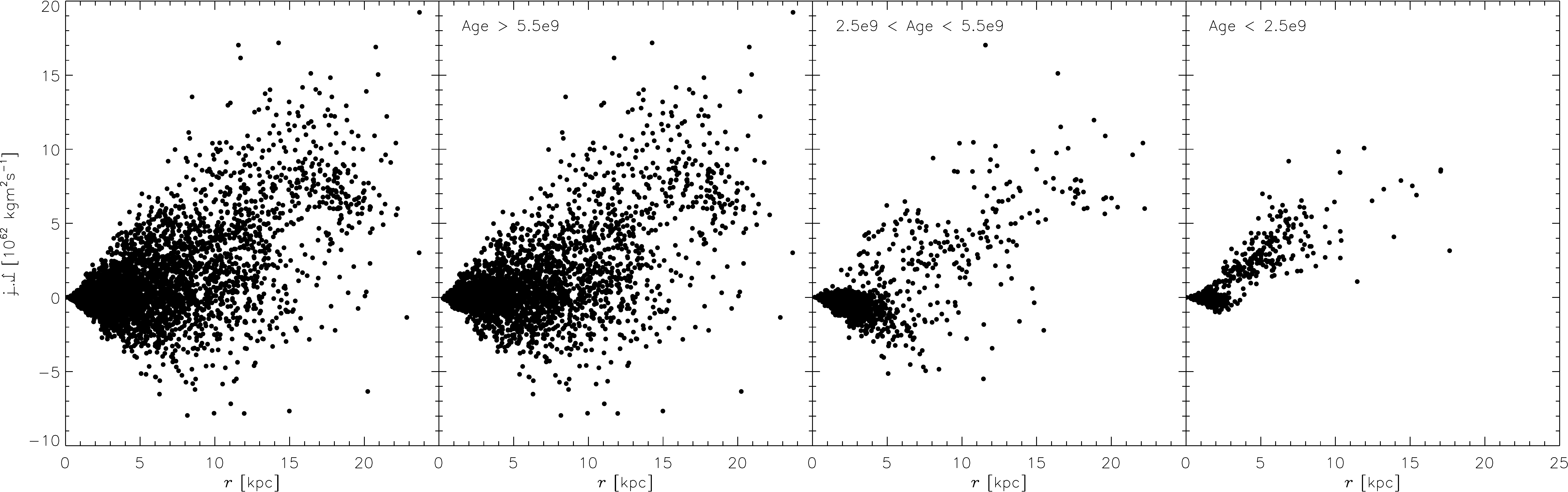}
	\caption{Component of angular momentum in the direction of the net angular momentum of the stars ($\mathbf{j}\boldsymbol{\cdot}\hat{\mathbf{J}}_*$) as a function of galactocentric distance for stars in ga0074 at $z=0$.
	The left-most panel shows all star particles, while the others bin stars by age.}
	\label{fig:ga0074_jr}
\end{figure*}

\begin{figure}
	\centering
	\includegraphics[width=0.48\textwidth,keepaspectratio]{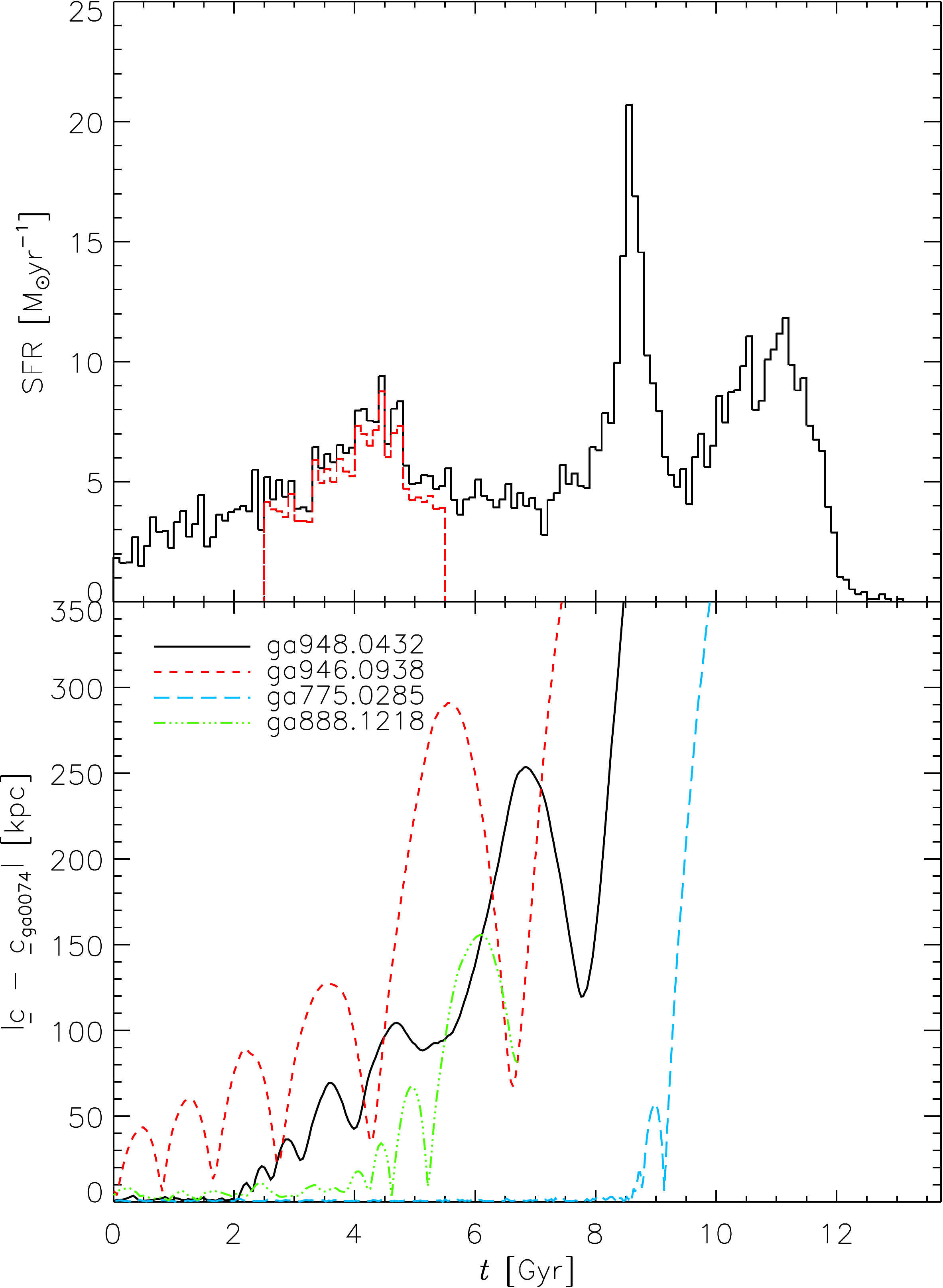}
	\caption{\emph{Top panel}: star formation history of ga0074.
	Solid black line is for the entire galaxy, dashed red line is for the stars of the KDC only.
	\emph{Lower panel}: separation of ga0074 and galaxies that merge onto it as a function of lookback time.}
	\label{fig:ga0074_sfh_sep}
\end{figure}

Galaxies ga0074 and ga0091 both display stellar KDCs at the present day, whereby the stars in the centre of the galaxies rotate in the opposite direction to both the gas and stars at the outskirts of the galaxies.
Their evolution initially parallels the evolution of the galaxies in Section \ref{sec:crgd}; both form in the field before falling into filaments, where they form a counter-rotating gas disc compared to the stellar motions by the same process as described in Section \ref{sec:crgd}.
The signature of counter-rotating gas discs begin to emerge at $z\sim0.25$ and $z\sim0.42$ in ga0074 and ga0091, respectively.
Subsequently, ga0074 and ga0091 both experience a merger (at $z\sim0.21$ and $z\sim 0.13$, respectively) with secondaries whose motion traces the bulk gas flows, and whose orbits are retrograde with respect the stellar angular momentum of the primary.
Fig. \ref{fig:orbits} shows the position of the secondary galaxy to each of ga0074 (left panel) and ga0091 (right panel) as it orbits and merges onto the primary.
The locations of star particles are shown, coloured by redshift, and the plane of the orbit is coincident with the plane of the page.

In both cases, the secondary galaxies have large orbital angular momentum and low mass compared to the primary.
The mergers onto ga0074 and ga0091 have stellar mass ratios of about 1:8 and 1:10, respectively.
This causes the accreted material to orbit further from the centre of the galaxy without disrupting the orbits of the pre-existing stars of the primary.
Since these accreted stars orbit in the same direction as the gas, and in the opposite direction to the central stars, the galaxies host a KDC following the mergers.

The stars constituting the KDC in ga0074 are relatively young, having formed between $2.5$ and $5.5$ Gyr before the present.
This is illustrated in Fig. \ref{fig:ga0074_jr} which shows the component of angular momentum in the direction of the net angular momentum  of stars as a function of galactrocentric distance at $z=0$.
The distribution of all particles (left-hand panel) is fairly uniform around $j_z=0$ at small radii, and skewed towards positive $j_z$ at large radii.
When only stars satisfying $2.5\,{\rm Gyr}<{\rm Age}<5.5\,{\rm Gyr}$ are considered (third panel), the KDC is clear, having negative $j_z$ and small radii ($r\ltsim5$ kpc).
The population of stars in that panel with large, positive $j_z$ formed in the accreted galaxy.

The top panel of Fig. \ref{fig:ga0074_sfh_sep} shows the star formation history of ga0074 (solid line) as well as just the stars that make up the KDC (red, dashed line).
The stars of the KDC are associated with a broad peak in star formation around 4 Gyr before present.
This star formation consumes much of the gas that co-rotates with it, which is why no kinematically decoupled gas kinematics are seen in this galaxy { at $z=0$} (Fig. \ref{fig:gaAll}).
It is not triggered by interactions with the secondary galaxy shown in Fig. \ref{fig:orbits}, but with a much less massive, gas-rich galaxy.

The lower panel of Fig. \ref{fig:ga0074_sfh_sep} shows the separation of ga0074 from galaxies that interact and merge with it, as a function of time.
The secondary galaxy shown in Fig. \ref{fig:orbits} is shown by the solid black line (ga948.0432).
ga888.1218 (dot-dashed green line) is a low mass, gas-rich system, with stellar mass fraction 0.09 and gas mass fraction 0.32\footnote{ The mass fraction of component X -- where X can be any of gas, stars, or dark matter -- of a galaxy is given by $M_{\rm X}/\left(M_{\rm gas}+M_*+M_{\rm DM}\right)$}.
The start of the star formation event that produces the stars of the KDC (red dashed line in upper panel of Fig. \ref{fig:ga0074_sfh_sep}) coincides with the merger of ga888.1218 onto ga0074.
Note that there is no peak in star formation associated with the merger of ga948.0432.

The evolution of ga0091 is similar, and slightly simpler.
Approximately 4.5 Gyr before the present, a counter rotating gas disc forms from the accretion of extra-galactic gas, causing the co-rotating gas previously present to fall to the centre of the galaxy and fuel low-level AGN activity (see also Section \ref{sec:crgd}).
About 1.7 Gyr before the present, the secondary galaxy shown in Fig. \ref{fig:orbits} merges with ga0091.
As for ga0074, the accreted stars orbit in the outskirts of the galaxy due to the large orbital angular momentum of the secondary, nearly co-planar with the gas.
The stars that formed in ga0091 (rather than being acquired through mergers) are mostly at the centre of the galaxy, and form the KDC.

In summary, there are three key stages to the formation of the KDCs.
These are: 1) the galaxies form in the field, and their gas and stars co-rotate; 2) the galaxies fall into a dark matter filament in which the bulk gas flow is such that a counter-rotating gas disc forms (as for the galaxies in Section \ref{sec:crgd}); 3) the minor merger of a low mass galaxy (whose orbit traces the large-scale gas flow) deposits high angular momentum stars in the outskirts of the galaxy, without disrupting the orbits of the pre-existing stars.
Note that the only difference between these galaxies and the galaxies with counter-rotating gas is that these galaxies subsequently experience a minor merger with a galaxy on a retrograde orbit.
Ultimately, this means that the formation of the KDCs depends on the relative orientation of the galaxy angular momentum and large-scale gas flows in the filaments.
With our SN and AGN feedback, additional star formation is not triggered by either the gas accretion from the filaments (Fig. \ref{fig:dm}) or by minor mergers at later times (Fig. \ref{fig:ga0074_sfh_sep}).

\subsection{Observability}\label{sec:observe}

{
Observed galaxy kinematics are derived from IFU data by fitting emission and absorption features, biasing the measurements in favour of the gas and stars that give rise to them.
Most IFUs operate in optical wavelengths, where strong emission lines like H$\alpha$, [O\,{\sc iii}], and [N\,{\sc ii}] are present.
These lines trace H\,{\sc ii} regions around young stars, as well as AGN activity.
In the galaxies considered in this paper, AGN activity is fairly low, and so we weight our gas kinematic measurements by SFR for a closer comparison to observations.
Stellar kinematics are weighted by $L_V$ since the $V$ band has significant overlap with the typical wavelength range of an IFU.
}

{
Fig. \ref{fig:weight} reproduces Fig. \ref{fig:gaAll}, but with kinematic maps now weighted as described above, rather than by mass.
Note also that the first column now shows SFR surface density, $\Sigma_{\rm SFR}$, rather than mass surface density for gas.
The range of values in each panel is kept the same as in Fig. \ref{fig:gaAll} for ease of comparison.
}

{
The gas kinematics of all galaxies are markedly different when weighted by SFR, suggesting that the ionised and atomic phases posses different kinematics.
This result highlights the need to examine all gas phases in observations of galaxies in order to properly understand their evolution \citep[see also][]{federrath17}.
However, obtaining kinematic maps from H\,{\sc i} observations is difficult for all but the closest galaxies \citep[e.g.,][]{kam17}.
}

{
For the stars, the KDCs of ga0074 and ga0091 are less clear when weighted by $L_V$, and the maps of $\phi$ more closely resemble their counterparts for gas.
$L_V$ is a better tracer of young stellar populations, which are more likely to have similar kinematics to the gas.
In ga0099, the direction of $\phi$ is reversed compared to the mass-weighted map in Fig. \ref{fig:gaAll}.
As for the gas, different kinematic populations are apparent when different weights are used, which may affect the interpretation of observations.
\citet{mcdermid06} showed that small-scale (10 - 100 pc) KDCs in fast rotators become harder to observe as the KDC population ages.
For the galaxies of Section \ref{sec:kdc}, there is no burst of star formation associated with the formation of the KDC, making it hard to see in the luminosity-weighted maps.
}

\section{Conclusions}\label{sec:conc}

We have analysed data from a cosmological hydrodynamical simulation that includes detailed models for star formation and stellar feedback, as well as BH formation and AGN feedback.
Our aim was to explain the formation of kinematically atypical galaxies from our simulation.
We demonstrated that projection effects do not change the distribution of angles between gas and stellar angular momenta (Fig. \ref{fig:costheta}).
Furthermore, of the 82 sufficiently well resolved galaxies, we identified two galaxies with kinematically distinct stellar cores (Section \ref{sec:kdc}) and three with counter-rotating gas discs (Section \ref{sec:crgd}).

{
All five of the galaxies studied formed in the field, later falling into a dark matter filament of the cosmic web.
The angular momentum of the galaxies and the direction of gas flow in the filaments are such that gas accreted from the filaments forms a counter-rotating disc within the galaxies (Section \ref{sec:crgd}).
Two of the galaxies subsequently experience a minor merger (stellar mass ratio $\sim1/10$) with a secondary on a retrograde orbit with respect to the angular momentum of the stars of the primary; no additional star formation is induced by these minor mergers (Fig. \ref{fig:ga0074_sfh_sep}).
The stars of the secondary galaxy are accreted onto the outskirts of the primary, where they rotate in the same direction to the gas, and opposite to the stars of the primary, giving rise to the KDC.
Overall, it is the relative orientation of flows within large-scale structure compared to the angular momentum of the primary galaxy that gives rise to the counter-rotating gas discs and KDCs.
}


Earlier studies have focused on the role of major mergers in creating KDCs.
In this work, we have examined well resolved, present-day galaxies that show atypical kinematics, and found that none forms due to major mergers.
The absence of KDCs formed by major mergers could indicate that: they are short lived; the simulation box is too small to provide a large enough sample of galaxies; or the resolution is not enough to see small-scale KDCs.
As cosmological simulations of ever greater volume and resolution are calculated, this situation will improve. 

Counter-rotating gas discs form due to prolonged accretion of gas from the cosmic web.
The gas exchanges angular momentum with the pre-existing, co-rotating gas in the galaxy, causing it to fall towards the galaxy centre.
This triggers low-level AGN feedback, which reduces the SFR in the galaxy, but not enough to move off the star formation main sequence (Fig. \ref{fig:sfh_jsjg}).
In the absence of AGN feedback, star formation could occur in the counter-rotating gas disc, which could lead to the formation of $2\sigma$ galaxies (Section \ref{sec:intro}).
This may happen at high redshift when the black holes are less massive, and AGN feedback is correspondingly weaker.
All three of the galaxies with counter-rotating gas at $z=0$ are embedded in dark matter filaments (Fig. \ref{fig:dm}), but away from the nodes in the cosmic web where clusters are found.
This allows for the accretion of large quantities of gas, while minimising the probability of experiencing a major merger that could destroy the gas disc.


\section*{Acknowledgements}
We thank the referee for their useful comments, which improved the clarity and quality of this paper.
PT thanks A.~Thomas and Y.~Jin for useful discussions related to Voronoi binning the simulated data.
C.F.~gratefully acknowledges funding provided by the Australian Research Council's Discovery Projects (grants~DP150104329 and~DP170100603) and by the Australia-Germany Joint Research Cooperation Scheme (UA-DAAD).
The simulations presented in this work used high performance computing resources provided by the Leibniz Rechenzentrum and the Gauss Centre for Supercomputing (grants pr32lo, pr48pi and GCS Large-scale project 10391), the Partnership for Advanced Computing in Europe (PRACE grant pr89mu), the Australian National Computational Infrastructure (grant ek9), and the Pawsey Supercomputing Centre with funding from the Australian Government and the Government of Western Australia, in the framework of the National Computational Merit Allocation Scheme and the ANU Allocation Scheme.
This work has made use of the University of Hertfordshire Science and Technology Research Institute high-performance computing facility.
Finally, we thank V.~Springel for providing {\sc GADGET-3}.


\bibliographystyle{mn2e}
\bibliography{/Users/ptaylor/papers/refs}

\begin{thebibliography}{}

\bibitem[\protect\citeauthoryear{{Agarwal}, {Khochfar}, {Johnson}, {Neistein},
  {Dalla Vecchia} \& {Livio}}{{Agarwal} et~al.}{2012}]{agarwal12}
{Agarwal} B.,  {Khochfar} S.,  {Johnson} J.~L.,  {Neistein} E.,  {Dalla
  Vecchia} C.,    {Livio} M.,  2012, \mnras, 425, 2854

\bibitem[\protect\citeauthoryear{{Algorry}, {Navarro}, {Abadi}, {Sales},
  {Steinmetz} \& {Piontek}}{{Algorry} et~al.}{2014}]{algorry14}
{Algorry} D.~G.,  {Navarro} J.~F.,  {Abadi} M.~G.,  {Sales} L.~V.,  {Steinmetz}
  M.,    {Piontek} F.,  2014, \mnras, 437, 3596

\bibitem[\protect\citeauthoryear{{Becerra}, {Greif}, {Springel} \&
  {Hernquist}}{{Becerra} et~al.}{2015}]{becerra15}
{Becerra} F.,  {Greif} T.~H.,  {Springel} V.,    {Hernquist} L.~E.,  2015,
  \mnras, 446, 2380

\bibitem[\protect\citeauthoryear{{Bland-Hawthorn}}{{Bland-Hawthorn}}{2015}]{Hector}
{Bland-Hawthorn} J.,  2015, in {Ziegler} B.~L.,  {Combes} F.,  {Dannerbauer}
  H.,   {Verdugo} M.,  eds,  IAU Symposium Vol. 309, Galaxies in 3D across the
  Universe. pp 21--28

\bibitem[\protect\citeauthoryear{{Bois} et~al.,}{{Bois} et~al.}{2011}]{bois11}
{Bois} M.  et~al., 2011, \mnras, 416, 1654

\bibitem[\protect\citeauthoryear{{Brodie} et~al.,}{{Brodie}
  et~al.}{2014}]{brodie14}
{Brodie} J.~P.  et~al., 2014, \apj, 796, 52

\bibitem[\protect\citeauthoryear{{Bromm}, {Coppi} \& {Larson}}{{Bromm}
  et~al.}{2002}]{bromm02}
{Bromm} V.,  {Coppi} P.~S.,    {Larson} R.~B.,  2002, \apj, 564, 23

\bibitem[\protect\citeauthoryear{{Bromm} \& {Loeb}}{{Bromm} \&
  {Loeb}}{2003}]{bromm03}
{Bromm} V.,  {Loeb} A.,  2003, \apj, 596, 34

\bibitem[\protect\citeauthoryear{{Cappellari}}{{Cappellari}}{2016}]{cappellari16}
{Cappellari} M.,  2016, \araa, 54, 597

\bibitem[\protect\citeauthoryear{{Cappellari} \& {Copin}}{{Cappellari} \&
  {Copin}}{2003}]{cappellari03}
{Cappellari} M.,  {Copin} Y.,  2003, \mnras, 342, 345

\bibitem[\protect\citeauthoryear{{Cappellari} et~al.,}{{Cappellari}
  et~al.}{2011}]{cappellari11}
{Cappellari} M.  et~al., 2011, \mnras, 416, 1680

\bibitem[\protect\citeauthoryear{{Coccato}, {Morelli}, {Corsini}, {Buson},
  {Pizzella}, {Vergani} \& {Bertola}}{{Coccato} et~al.}{2011}]{coccato11}
{Coccato} L.,  {Morelli} L.,  {Corsini} E.~M.,  {Buson} L.,  {Pizzella} A.,
  {Vergani} D.,    {Bertola} F.,  2011, \mnras, 412, L113

\bibitem[\protect\citeauthoryear{{Croom} et~al.,}{{Croom}
  et~al.}{2012}]{croom12}
{Croom} S.~M.  et~al., 2012, \mnras, 421, 872

\bibitem[\protect\citeauthoryear{{Davis} et~al.,}{{Davis}
  et~al.}{2011}]{davis11}
{Davis} T.~A.  et~al., 2011, \mnras, 417, 882

\bibitem[\protect\citeauthoryear{{de Zeeuw} et~al.,}{{de Zeeuw}
  et~al.}{2002}]{dezeeuw02}
{de Zeeuw} P.~T.  et~al., 2002, \mnras, 329, 513

\bibitem[\protect\citeauthoryear{{Dopita} et~al.,}{{Dopita}
  et~al.}{2015}]{dopita15}
{Dopita} M.~A.  et~al., 2015, \apjs, 217, 12

\bibitem[\protect\citeauthoryear{{Dubois}, {Gavazzi}, {Peirani} \&
  {Silk}}{{Dubois} et~al.}{2013}]{dubois13}
{Dubois} Y.,  {Gavazzi} R.,  {Peirani} S.,    {Silk} J.,  2013, \mnras, 433,
  3297

\bibitem[\protect\citeauthoryear{{Emsellem} et~al.,}{{Emsellem}
  et~al.}{2007}]{emsellem07}
{Emsellem} E.  et~al., 2007, \mnras, 379, 401

\bibitem[\protect\citeauthoryear{{Federrath} et~al.,}{{Federrath}
  et~al.}{2017}]{federrath17}
{Federrath} C.  et~al., 2017, \mnras, 468, 3965

\bibitem[\protect\citeauthoryear{{Haardt} \& {Madau}}{{Haardt} \&
  {Madau}}{1996}]{haardt96}
{Haardt} F.,  {Madau} P.,  1996, \apj, 461, 20

\bibitem[\protect\citeauthoryear{{Hinshaw} et~al.,}{{Hinshaw}
  et~al.}{2013}]{wmap9}
{Hinshaw} G.  et~al., 2013, \apjs, 208, 19

\bibitem[\protect\citeauthoryear{{Hoffman}, {Cox}, {Dutta} \&
  {Hernquist}}{{Hoffman} et~al.}{2010}]{hoffman10}
{Hoffman} L.,  {Cox} T.~J.,  {Dutta} S.,    {Hernquist} L.,  2010, \apj, 723,
  818

\bibitem[\protect\citeauthoryear{{Hosokawa}, {Hirano}, {Kuiper}, {Yorke},
  {Omukai} \& {Yoshida}}{{Hosokawa} et~al.}{2016}]{hosokawa16}
{Hosokawa} T.,  {Hirano} S.,  {Kuiper} R.,  {Yorke} H.~W.,  {Omukai} K.,
  {Yoshida} N.,  2016, \apj, 824, 119

\bibitem[\protect\citeauthoryear{{Jesseit}, {Naab}, {Peletier} \&
  {Burkert}}{{Jesseit} et~al.}{2007}]{jesseit07}
{Jesseit} R.,  {Naab} T.,  {Peletier} R.~F.,    {Burkert} A.,  2007, \mnras,
  376, 997

\bibitem[\protect\citeauthoryear{{Jin} et~al.,}{{Jin} et~al.}{2016}]{jin16}
{Jin} Y.  et~al., 2016, \mnras, 463, 913

\bibitem[\protect\citeauthoryear{{Kam}, {Carignan}, {Chemin}, {Foster}, {Elson}
  \& {Jarrett}}{{Kam} et~al.}{2017}]{kam17}
{Kam} S.~Z.,  {Carignan} C.,  {Chemin} L.,  {Foster} T.,  {Elson} E.,
  {Jarrett} T.~H.,  2017, \aj, 154, 41

\bibitem[\protect\citeauthoryear{{Khochfar} et~al.,}{{Khochfar}
  et~al.}{2011}]{khochfar11}
{Khochfar} S.  et~al., 2011, \mnras, 417, 845

\bibitem[\protect\citeauthoryear{{Kobayashi}}{{Kobayashi}}{2004}]{ck04}
{Kobayashi} C.,  2004, \mnras, 347, 740

\bibitem[\protect\citeauthoryear{{Kobayashi}, {Karakas} \& {Umeda}}{{Kobayashi}
  et~al.}{2011}]{ck11b}
{Kobayashi} C.,  {Karakas} A.~I.,    {Umeda} H.,  2011, \mnras, 414, 3231

\bibitem[\protect\citeauthoryear{{Kobayashi} \& {Nakasato}}{{Kobayashi} \&
  {Nakasato}}{2011}]{ck11a}
{Kobayashi} C.,  {Nakasato} N.,  2011, \apj, 729, 16

\bibitem[\protect\citeauthoryear{{Kobayashi} \& {Nomoto}}{{Kobayashi} \&
  {Nomoto}}{2009}]{ck09}
{Kobayashi} C.,  {Nomoto} K.,  2009, \apj, 707, 1466

\bibitem[\protect\citeauthoryear{{Kobayashi}, {Springel} \&
  {White}}{{Kobayashi} et~al.}{2007}]{ck07}
{Kobayashi} C.,  {Springel} V.,    {White} S.~D.~M.,  2007, \mnras, 376, 1465

\bibitem[\protect\citeauthoryear{{Kobayashi}, {Umeda}, {Nomoto}, {Tominaga} \&
  {Ohkubo}}{{Kobayashi} et~al.}{2006}]{ck06}
{Kobayashi} C.,  {Umeda} H.,  {Nomoto} K.,  {Tominaga} N.,    {Ohkubo} T.,
  2006, \apj, 653, 1145

\bibitem[\protect\citeauthoryear{{Koushiappas}, {Bullock} \&
  {Dekel}}{{Koushiappas} et~al.}{2004}]{koushiappas04}
{Koushiappas} S.~M.,  {Bullock} J.~S.,    {Dekel} A.,  2004, \mnras, 354, 292

\bibitem[\protect\citeauthoryear{{Krajnovi{\'c}} et~al.,}{{Krajnovi{\'c}}
  et~al.}{2008}]{krajnovic08}
{Krajnovi{\'c}} D.  et~al., 2008, \mnras, 390, 93

\bibitem[\protect\citeauthoryear{{Krajnovi{\'c}} et~al.,}{{Krajnovi{\'c}}
  et~al.}{2011}]{krajnovic11}
{Krajnovi{\'c}} D.  et~al., 2011, \mnras, 414, 2923

\bibitem[\protect\citeauthoryear{{Kroupa}}{{Kroupa}}{2008}]{kroupa08}
{Kroupa} P.,  2008, in {Knapen} J.~H.,  {Mahoney} T.~J.,   {Vazdekis} A.,  eds,
   Astronomical Society of the Pacific Conference Series Vol. 390, Pathways
  Through an Eclectic Universe. p.~3

\bibitem[\protect\citeauthoryear{{Ma}, {Greene}, {McConnell}, {Janish},
  {Blakeslee}, {Thomas} \& {Murphy}}{{Ma} et~al.}{2014}]{ma14}
{Ma} C.-P.,  {Greene} J.~E.,  {McConnell} N.,  {Janish} R.,  {Blakeslee} J.~P.,
   {Thomas} J.,    {Murphy} J.~D.,  2014, \apj, 795, 158

\bibitem[\protect\citeauthoryear{{Madau} \& {Rees}}{{Madau} \&
  {Rees}}{2001}]{madau01}
{Madau} P.,  {Rees} M.~J.,  2001, \apjl, 551, L27

\bibitem[\protect\citeauthoryear{{McDermid} et~al.,}{{McDermid}
  et~al.}{2006}]{mcdermid06}
{McDermid} R.~M.  et~al., 2006, \mnras, 373, 906

\bibitem[\protect\citeauthoryear{{Naab} et~al.,}{{Naab} et~al.}{2014}]{naab14}
{Naab} T.  et~al., 2014, \mnras, 444, 3357

\bibitem[\protect\citeauthoryear{{Osman} \& {Bekki}}{{Osman} \&
  {Bekki}}{2017}]{osman17}
{Osman} O.,  {Bekki} K.,  2017, \mnras, 471, L87

\bibitem[\protect\citeauthoryear{{Regan}, {Johansson} \& {Wise}}{{Regan}
  et~al.}{2016}]{regan16a}
{Regan} J.~A.,  {Johansson} P.~H.,    {Wise} J.~H.,  2016, \mnras, 459, 3377

\bibitem[\protect\citeauthoryear{{S{\'a}nchez} et~al.,}{{S{\'a}nchez}
  et~al.}{2012}]{sanchez12}
{S{\'a}nchez} S.~F.  et~al., 2012, \aap, 538, A8

\bibitem[\protect\citeauthoryear{{Schneider}, {Ferrara}, {Natarajan} \&
  {Omukai}}{{Schneider} et~al.}{2002}]{schneider02}
{Schneider} R.,  {Ferrara} A.,  {Natarajan} P.,    {Omukai} K.,  2002, \apj,
  571, 30

\bibitem[\protect\citeauthoryear{{Springel}}{{Springel}}{2005}]{springel05gadget}
{Springel} V.,  2005, \mnras, 364, 1105

\bibitem[\protect\citeauthoryear{{Sutherland} \& {Dopita}}{{Sutherland} \&
  {Dopita}}{1993}]{sutherland93}
{Sutherland} R.~S.,  {Dopita} M.~A.,  1993, \apjs, 88, 253

\bibitem[\protect\citeauthoryear{{Taylor}, {Federrath} \& {Kobayashi}}{{Taylor}
  et~al.}{2017}]{pt17a}
{Taylor} P.,  {Federrath} C.,    {Kobayashi} C.,  2017, \mnras, 469, 4249

\bibitem[\protect\citeauthoryear{{Taylor} \& {Kobayashi}}{{Taylor} \&
  {Kobayashi}}{2014}]{pt14}
{Taylor} P.,  {Kobayashi} C.,  2014, \mnras, 442, 2751

\bibitem[\protect\citeauthoryear{{Taylor} \& {Kobayashi}}{{Taylor} \&
  {Kobayashi}}{2015a}]{pt15a}
{Taylor} P.,  {Kobayashi} C.,  2015a, \mnras, 448, 1835

\bibitem[\protect\citeauthoryear{{Taylor} \& {Kobayashi}}{{Taylor} \&
  {Kobayashi}}{2015b}]{pt15b}
{Taylor} P.,  {Kobayashi} C.,  2015b, \mnras, 452, L59

\bibitem[\protect\citeauthoryear{{Taylor} \& {Kobayashi}}{{Taylor} \&
  {Kobayashi}}{2016}]{pt16}
{Taylor} P.,  {Kobayashi} C.,  2016, \mnras, 463, 2465

\bibitem[\protect\citeauthoryear{{Taylor} \& {Kobayashi}}{{Taylor} \&
  {Kobayashi}}{2017}]{pt17b}
{Taylor} P.,  {Kobayashi} C.,  2017, \mnras, 471, 3856

\bibitem[\protect\citeauthoryear{{Thomas} et~al.,}{{Thomas}
  et~al.}{2017}]{thomas17}
{Thomas} A.~D.  et~al., 2017, \apjs, 232, 11

\bibitem[\protect\citeauthoryear{{Yan} et~al.,}{{Yan} et~al.}{2016}]{yan16}
{Yan} R.  et~al., 2016, \aj, 152, 197

\end{thebibliography}


\appendix
\section{Kinematic maps weighted by $L_V$ and SFR}\label{sec:app:weight}

\begin{figure*}
	\centering
	\includegraphics[totalheight=0.95\textheight,keepaspectratio]{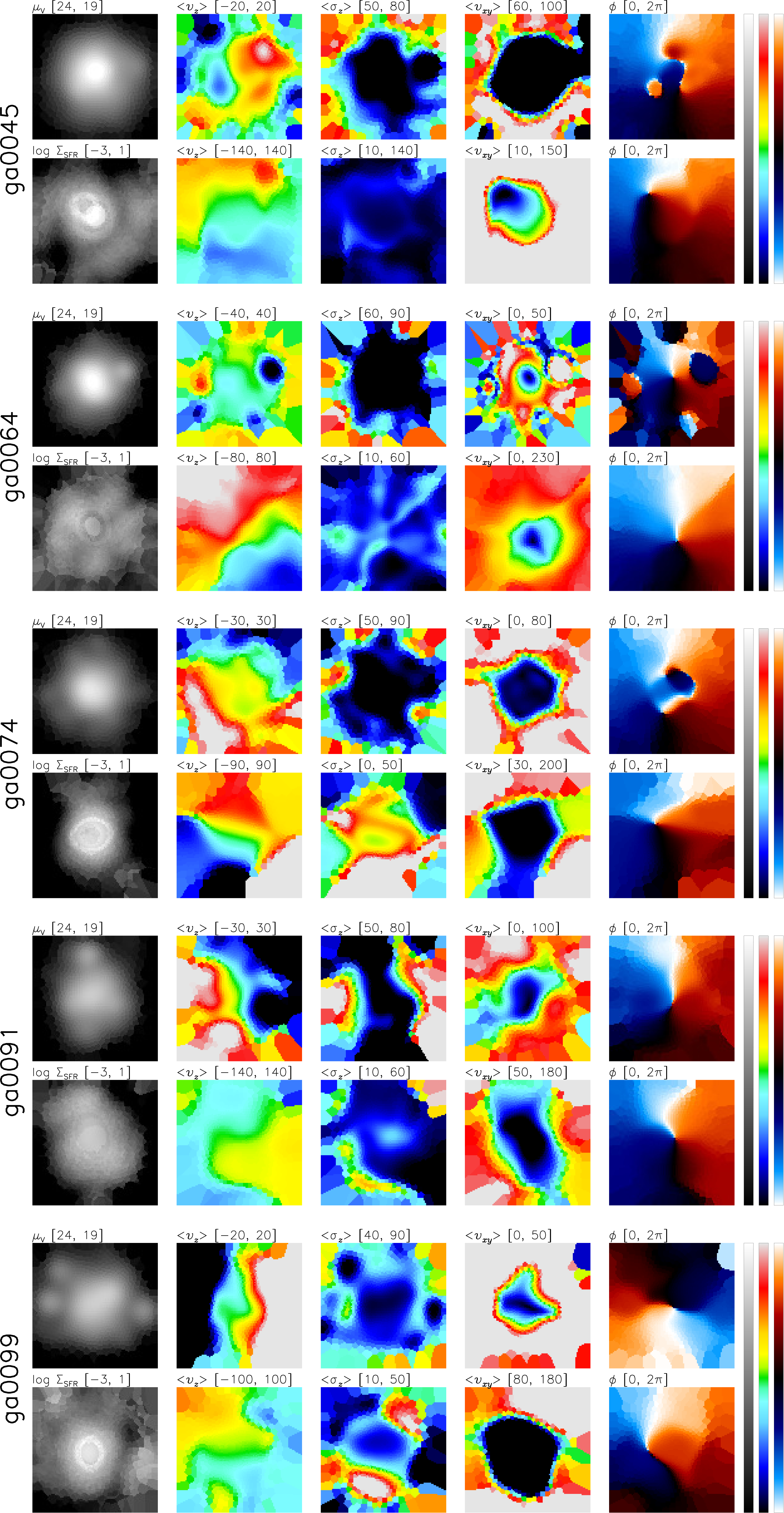}
	\caption{Kinematic maps for the galaxies shown in Fig. \ref{fig:gaAll}, with stellar velocities weighted by $L_V$, and gas velocities by SFR.
	The upper row for each galaxy shows stellar properties, and the lower row is for gas.
	The first column shows $V$-band surface brightness (mag\,arcsec$^{-2}$) and SFR surface density (M$_\odot$\,yr$^{-1}$\,kpc$^{-2}$), the second $\left<v_z\right>$ (km\,s$^{-1}$), the third $\sigma_z$ (km\,s$^{-1}$), the fourth $v_{xy}$ (km\,s$^{-1}$), and the fifth $\phi$.
	The range of values in each panel is shown in brackets.}
	\label{fig:weight}
\end{figure*}

Fig. \ref{fig:weight} reproduces Fig. \ref{fig:gaAll}, but with kinematic maps now weighted as described above, rather than by mass.
Note also that the first column now shows SFR surface density, $\Sigma_{\rm SFR}$, rather than mass surface density for gas.
The range of values in each panel is kept the same as in Fig. \ref{fig:gaAll} for ease of comparison.

\section{Kinematic maps with differing target S/N}\label{sec:app:sn}

\begin{figure*}
	\centering
	\includegraphics[totalheight=0.95\textheight,keepaspectratio]{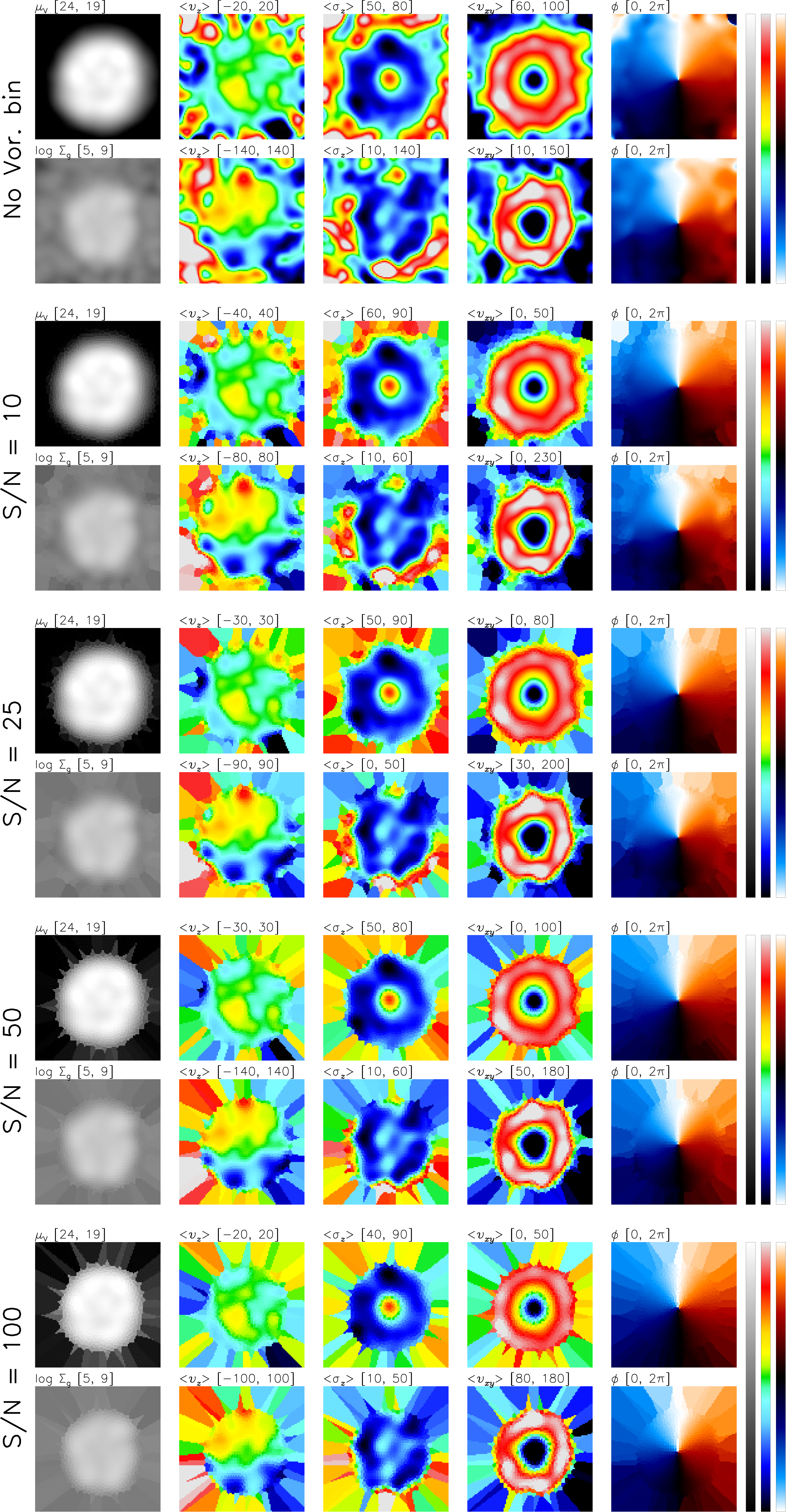}
	\caption{Kinematic maps of ga0008, Voronoi binned to different target S/N, but otherwise the same as Fig. \ref{fig:ga0008}.
	The range of values in each panel is shown in brackets.}
	\label{fig:sn}
\end{figure*}

Fig. \ref{fig:sn}  shows the kinematic maps of galaxy ga0008 as in Fig. \ref{fig:ga0008}, but Voronoi binned to different target S/N.
The first two rows also show the un-binned images.
The binning mainly affects the outskirts of the galaxy, where the surface brightness is lower, but for all values of S/N the same qualitative and quantitative features are seen within the galaxy.

\bsp

\label{lastpage}

\end{document}